\documentclass{article}

\usepackage{epsfig} 
\usepackage{amsmath}
\usepackage{amsfonts}
\usepackage{graphicx}

 \usepackage{amssymb}

\date{\today} 
 
\newcommand{\bnabla}{\mbox{\boldmath $\nabla$}}
\newcommand{\bOmega}{\mbox{\boldmath $\Omega$}}

\newcommand{\insertplot}[5]{\begin{figure}
 \hfill\hbox to 0.05in{\vbox to #5in{\vfill
 \inputplot{#1}{#4}{#5}}\hfill}
 \hfill\vspace{-.1in}
 \caption{#2}\label{#3}
 \end{figure}}
 \newcommand{\inputplot}[3]{
 \special{ps: plotfile #1}
}
\newcounter{fig}

\newcommand{\la}{\lambda}

\newcommand{\f}{\phi}
\newcommand{\F}{\Phi}

\newcommand{\ka}{\kappa}
\newcommand{\al}{\alpha}

\newcommand{\Ga}{\Gamma}

\newcommand{\si}{\sigma}
\newcommand{\Si}{\Sigma}

\newcommand{\ee}{\end{equation}}
\newcommand{\eea}{\end{eqnarray}}
\newcommand{\be}{\begin{equation}}
\newcommand{\bea}{\begin{eqnarray}}

\newcommand{\pa}{\partial}

\newcommand{\vep}{\varepsilon}

\def\theequation{\arabic{equation}}

\newcommand{\re}[1]{(\ref{#1})}
\newcommand{\R}{{\rm I \hspace{-0.52ex} R}}

\newcommand{\eins}{1\hspace{-0.56ex}{\rm I}}

\begin{document}

\title{ 
Instantonic dyons of Yang-Mills--Chern-Simons models
\\
in $d=2n+1$ dimensions, $n>2$
} 
 \vspace{1.5truecm}
\author{ 
{\large Eugen Radu}$^{\dagger}$ 
and {\large D. H. Tchrakian}$^{\star \diamond }$ 
\vspace*{0.2cm}
\\
$^{\dagger}${\small Institut f\"ur Physik, Universit\"at Oldenburg,
D-26111 Oldenburg, Germany} 
   \\
$^{\star}${\small 
School of Theoretical Physics -- DIAS, 10 Burlington
Road, Dublin 4, Ireland} \\
$^{\diamond}${\small  Department of Computer Science,
National University of Ireland Maynooth,
Maynooth,
Ireland}}

\def\theequation{\thesection.\arabic{equation}}

\maketitle

\begin{abstract}
We investigate finite energy solutions 
of Yang-Mills--Chern-Simons systems in odd spacetime dimensions, $d=2n+1$, with $n>2$.
This can be carried out systematically for all $n$, but
the cases $n=3,4$ corresponding to a $7,8$ dimensional spacetime are treated concretely.
These are static and spherically symmetric configurations, defined in a flat Minkowski background. 
The value of the electric charge is fixed by the Chern-Simons
coupling constant. 
\end{abstract}

\section{Introduction}
Yang-Mills (YM) fields in $d=D+1$ dimensional spacetime have interesting properties,
especially in the static limit where 
the solutions to the equations of motion describe finite energy topologically stable solutions.
Famously, in the case $D=3$ these
are the 't~Hooft-Polyakov monopoles~\cite{'tHooft:1974qc,Polyakov:1974ek}
on $\R^3$ whose topological charge is the magnetic flux. In fact, monopoles exist also
on all $\R^D$~\cite{Tchrakian:2010ar}. Monopoles are supported by
Yang--Mills--Higgs (YMH) systems, where the Higgs field (which is always
an iso-$D-$vector) defines the topology completely by
virtue of the requirement of finiteness of energy ~\cite{Arafune:1974uy,Tchrakian:2002ti}. 

However, in $d=3+1$ dimensional spacetime,
there exists also the Julia-Zee dyon~\cite{Julia:1975ff}
which in addition to magnetic flux also possesses an electric flux. Unlike the magnetic charge,
the electric charge of a dyon, while it is a global charge, is not a
topological charge. 
The topological charge appearing in the Julia--Zee (JZ) dyon is the magnetic monopole
charge given by the spacelike component $A_i$ of the gauge connection (and the Higgs field $\F$),
while the electric flux is given by the timelike component $A_0$ (and $\F$). By construction,
$A_0$ and $\F$ are proportional in the absence of a Higgs potential.
Now  the existence of the Julia--Zee (JZ) dyon is predicated on the presence of a monopole,
as well as the availability of the 't~Hooft electromagnetic tensor. This mechanism is restricted to
$3+1$ dimensional spacetime. Although monopoles can be constructed in all dimensions, the definition
of a 't~Hooft tensor in higher dimensional space is problematic, as discussed in
\cite{Tchrakian:2010ar}. So it is not open to us to exploit the $D-$dimensional monopoles for
the purpose of constructing dyons in higher dimensions.

To the best of our knowledge, the only dyon in higher
dimensions is the dyonic instanton of Lambert and Tong~\cite{Lambert:1999ua} (LT) in $4+1$
dimensional spacetime. The topological charge of the LT dyon is
the usual instanton number on the $\R^4$ subspace, described by the
spacelike component $A_i$ of the gauge connection. The electic flux is again given by the
timelike component $A_0$ and the Higgs field $\F$, as for the JZ dyon. Unlike in the case of the latter however,
$A_0$ of the LT dyon is actually equal to $\F$ and not just proportional to it. The dyonic instanton describes
a global electric flux in addition to
the Pontryagin charge (of the second Chern class), the latter being the topological charge analogous to
the magnetic charge of the Julia-Zee dyon.

Our aim in the present work is to construct a new type  of instantonic dyons 
in (higher) odd dimensional
spacetimes. For these solutions, the spacelike sector is stabilised by a topological charge, and which support
also an electric flux. As in the case of the LT dyonic instanton, where the ``magnetic charge'' is the $2-$nd
Chern-Pontryagin charge, the topological charge here is 
the $n-$th Chern-Pontryagin charge on $\R^D$, $D=2n$. The latter is given by the
spacelike, or ``magnetic'', component $A_i$ of the gauge connection.
Our solutions share this property with the LT dyonic instanton.
Naturally, in our case too, the electric flux associated with the solutions proposed is given by the
timelike, or ``electric'', component $A_0$.

The models proposed here differ from those of JZ~\cite{Julia:1975ff} and of
LT~\cite{Lambert:1999ua} in two respects. In common with these, 
they feature non-Abelian Yang-Mills fields, but in
contrast they contain no Higgs field. The Higgs field is employed to support the
magnetic (topological) charge of the JZ dyon, and in the case of the
LT dyonic instanton, it supports the component $A_0$. The role corresponding
to the Higgs (kinetic) term is here 
played by the non-Abelian Chern-Simons (CS) density. Employing a CS term in
the action for the purpose of supporting $A_0$ is
standard, both for Abelian~\cite{Hong:1990yh,Jackiw:1990aw} and non-Abelian~\cite{NavarroLerida:2008uj}
systems in $2+1$ dimensions, as well as in higher dimensions~\cite{Brihaye:2009cc}. 

It is important to emphasise the difference in using the $instanton$ number as the
topological charge, rather than the $monopole$ charge. The connection $A_i$ of the instanton
behaves asymptotically as a {\it pure-gauge}, in contrast to
that of monopole which behaves as {\it half pure-gauge}. In
other words the latter is a Dirac-Yang~\cite{Dirac:1931kp,Yang:1977qv,Tchrakian:2008zz} monopole decaying as
$r^{-1}$, more slowly than the instanton. As a consequence, the energy density functionals of
monopoles in higher (than three spacelike)
dimensions cannot involve the usual quadratic Yang-Mills term whose energy integral
diverges. This is not the case for the faster (pure-gauge) decay of the instanton,
which enables the retention of a quadratic YM density in all dimensions,
with converging ``energy integral''. From a physical standpoint, this alone could be
considered a motivation for employing instanton numbers in preference to
monopole charges as topologiccal charges in higher dimensions.

The models we will introduce will feature the Yang-Mills sector,
consisting of some or all possible terms of the Yang-Mills
hierarchy~\cite{Tchrakian:1984gq,Tchrakian:1992ts}
(to be introduced below), and the Chern-Simons density in the appropriate
dimension. The topological charge is the $instanton$ number of
the given YM system on the ``spacelike'' subspace $\R^{D}$ (or $\R^{2n}$), which always includes the usual
(quadratic) Yang-Mills density, as well as at least one other member of the YM
hierarchy which is of sufficiently high order in the YM curvature
to enable the Derrick scaling requirement for the convergence of the ``energy integral''
to be satisfied. The application of the Derrick scaling requirement
Yang-Mills sector (on a Euclidean space) is rigorous.

 It is important to distinguish the status of the Derrick scaling requirement
in the case of dyons where the electric YM connection $A_0$ is introduced
in the covariant derivative of the Higgs field
as is the case for the JZ and
LT dyons, and in our case when it is introduced $via$ a CS term in the action
as was done in the case of the gravitating YMCS system in $4+1$ and higher dimensions~\cite{Brihaye:2009cc,Brihaye:2011nr}.
(Note also the Abelian~\cite{Hong:1990yh,Jackiw:1990aw}
and non-Abelian~\cite{NavarroLerida:2008uj} CS-Higgs
vortices in $2+1$ dimensions, where $A_0$ enters both in the Higgs covariant derivative and the the CS density.)
The status of Derrick scaling is subtle when there is a CS term in the action, and the non-Abelian case is
less transparent than the Abelian. In the Abelian case, the electric
gauge connection $A_0$ can be found explicitly by solving the Gauss Law equation, and substituting it into
the static energy density functional exposes the (Derrick) scaling properties.
In the non-Abelian cases \cite{NavarroLerida:2008uj,Brihaye:2009cc,Brihaye:2011nr} 
by contrast, it is not practicable to solve the Gauss Law equation to yield the (non-Abelian)
electric gauge connection $A_0$ explicitly. In those cases the solution of the Gauss Law equation
is implicit in the numerical process, resulting nonetheless in the required scaling contribution of the CS term
appearing in the action, now in the static energy density. Thus the CS density acts as a higher order term
enabling the Derrick scaling requirement for the finiteness of energy to be satisfied. 

Our formulation is in principle for all spacetime dimensions $d=2n+1$, $n\ge 3$.
We have considered spherically symmetric static solutions and the asymptotic
analysis for these is given for arbitrary $D=2n$.
Explicit, numerical, constructions however are restricted to the $n=3,4$ cases.

In Section {\bf 2} the models are introduced, and the imposition of spherical
symmetry as well as the resulting field equations are given in Section {\bf 3}.
The solutions are presented in Section {\bf 4}, which includes the asymptotic
analysis in the general case, as well as the concrete numerical construction
for the dyons in $7$ and $9$ dimensional spacetimes. In addition, we supply
three Appendices. Appendix {\bf A} presents a simplification of the (dynamical)
Chern-Simons density in the static limit. This simplifies the algebraic
calculations considerably, in particular enabling the formulation of the
arbitrary dimensional case. Appendix {\bf B}
gives the spherically symmetric Ansatz for the full $SO(D+2)$ Yang-Mills system. Appendix {\bf C} gives an
attempt at generalising the Lambert-Tong~\cite{Lambert:1999ua} dyonic instanton in $d=4+1$ to $d=4p+1$. While
this attempt is not completely satisfactory, the review of that material exposes the essential
difference of our instantonic dyons in $7,9,\dots$ dimensions and the LT dyonic instanton in $5$ dimensions.

\section{The models}

In selecting the models in $d=D+1$, with $D=2n$, we have invoked two criteria,
that of topological stability of the ``magnetic'' sector, and, that of
the existence of a global electric charge. Our first criterion is that the
spacelike components of the non-Abelian gauge fields describe a topologically stable (static)
field configuration charcterised by an instanton charge.
(The usual action integral here is interpreted as the ``energy integral''.) 
For these
configurations to have finite energy, it is necessary to employ at least two appropriately
scaling members of the Yang-Mills
hierarchy~\cite{Tchrakian:1984gq,Tchrakian:1992ts,Tchrakian:2010ar} on $\R^D=\R^{2n}$
\be
\label{pYM}
{\cal L}_{\rm YM}^{(p)}=\frac{1}{2\cdot(2p)!}\mbox{Tr}\,\bigg\{ F(2p)^2\bigg\}\,,
\ee
in terms of the ``magnetic'' field $A_i$. Later, when we introduce the ``electric'' field $A_0$, this definition
will be retained, but then on $d=D+1$ dimensional spacetime with Minkowskian signature. In \re{pYM},
$F(2p)$ is the totally antisymmetrised $p-$fold product of the YM curvature $2-$form $F(2)$.
We shall refer to the system \re{pYM} as a $p$-YM system, the $1$-YM system being the usual YM
density. Note that the $p$-YM density scales as $L^{-4p}$.
In any given spacetime dimension $d$, the constraint of antisymmetry of $F(2p)$
requires that the highest order curvature term $F(2P)$ is that with $d=2P$, which scales as $L^{-4P}$. 

Subject to this constraint, the most general YM system is the superposition of terms \re{pYM}
\be
\label{pYMsuperposn}
{\cal L}_{\rm YM}=\sum_{p=1}^{P}\frac{1}{2\cdot(2p)!}\mbox{Tr}\,\bigg\{ F(2p)^2\bigg\}\,,
\ee
the  $p=1$ term being the usual YM system.
Finiteness of the ``energy'' requires that Derrick scaling be satisfied, and given that the lowest order curvature
term with $p=1$ scales as $L^{-4}$ and the highest with $p=P$ as $L^{-2(D+1)}$, this is (more than) sufficient
for Derrick scaling to be satisfied on $\R^D$.
A subsystem of the superposition \re{pYMsuperposn} will be adopted as the Yang-Mills sector of
our models. 

Our choice of the Yang-Mills sector is made such that the solutions on $\R^{D}$, $i.e.$ the ``magnetic''
field configurations, have finite ``energy'' and be topologically stable, as is the case for the Julia-Zee dyon and
the dyonic instanton. The
topological stability stems from the following sets of inequalities
\be
\label{topineq1}
\mbox{Tr}\,  \bigg\{ F(2p_1)-\left(^{\star}F(2p_2)\right)(2p_1)  \bigg\}^2\ge 0\,,
\ee
where $^{\star}F(2p_2)$ is the Hodge dual of $F(2p_2)$, which is of course a $2p_1$-form, provided that the $(p_1,p_2)$
pair is a partition of $2(p_1+p_2)=D=2n$. This means that topological stability constrains the highest order curvature term in
\re{pYMsuperposn}, $\mbox{Tr} \bigg\{F(2P)^2\bigg\}$ be the $P=D$ term, rather than $P=D+1$ in Minkowski space.

It follows that for any $D=2n$, one has the ``energy'' lower bound
\be
\label{topineq}
{\cal L}_{\rm YM}^{(p_1,p_2)}\stackrel{\rm def.}=
\tau_1\,{\cal L}_{\rm YM}^{(p_1)}+\tau_2\,{\cal L}_{\rm YM}^{(p_2)}\ge{\cal C}_{(n)}\,,
\ee
for any partition~\footnote{When $p_1=p_2=p$,
with $D=4p$, the topological inequalities  \re{topineq1} and \re{topineq} are saturated and there result
self-dual BPST~\cite{Belavin:1975fg} configurations~\cite{Tchrakian:1984gq}, the $p=1$ case
being the usual BPST instanton. We have eschewed this choice here since no solutions exist for such systems when
the Chern-Simons terms are introduced. A discussion of this will be given in Section {\bf 5}.}
$n=p_1+p_2$. Here, ${\cal C}_{(n)}$ is the $n-$th Chern--Pontryagin density, and $(\tau_1,\tau_2)$ are
dimensionful (coupling) constants.

The criterion of topological stability \re{topineq} requires the presence of at least two YM terms.
Thus, for simplicity we shall restrict our definitions to systems consisting of exactly two $p$-YM terms.
Moreover for ``physical'' reasons, we will fix $p_1$ to $p_1=1$, so as to retain the usual (quadratic) $1$-YM term.
This in turn fixes fix $p_2$ to $p_2=n-1$, finally fixing the YM system to the superposition of the $1$-YM and 
the $(n-1)$-YM terms.
\bea
\label{YM1n-1}
{\cal L}_{\rm YM}&=&\tau_1\,{\cal L}_{\rm YM}^{(1)}+\tau_2\,{\cal L}_{\rm YM}^{(n-1)}\,.
\eea
Finally, the existence of 'instantons' of these systems requires that the YM connection takes its values in the chiral
representation of $SO(D)=SO(2n)$, such that the gauge group $G$ $must$ contain the subgroup $SO(D)$.
This completes the definition of the Yang-Mills sector.

Concerning the introduction of the Chern-Simons (CS) term   (which
fulfills our second criterion, namely that of supporting an electric field,)
 this is uniquely fixed by the dimension $D+1$ of the spacetime, and is
accompanied with the introduction of the ``electric'' connection $A_0$ to the whole system. The smallest simple gauge
group that supports a nonvanishing CS density is $G=SO(D+2)$, which finally fixes our choice of gauge group. In practice
the gauge group will be truncated to $G=SO(D)\times SO(2)$ because in the spherically symmetric case studied here,
it transpires from the asyptotic analysis in Section {\bf 4} that only $G=SO(D)\times SO(2)$ solutions can be found. (We expect that
in the presence of a negative cosmological constant there exist full $SO(D+2)$ solutions, as was found in \cite{Brihaye:2009cc} for $d=5$. It is
for this reason that the full $SO(D+2)$ spherically symmetric Ansatz is given in Appendix {\bf B}.) 

As in all dyonic systems,
the topological stability of the purely ``magnetic'' sector does not any more guarantee the existence of finite energy solutions.
(Indeed analytic proofs of existence for dyons are problematic and their existence is usually established numerically, as for the
JZ dyon (in the presence of a Higgs potential), or by explicit construction in closed form as in the case of the LT dyonic instanton.)

Therefore the Lagrangians we adopt are
\be
\label{YMCS}
{\cal L}_{\rm YMCS}={\cal L}_{\rm YM}+\ka\,{\cal L}_{\rm CS}^{(n)}\,,
\ee
in which ${\cal L}_{\rm YM}$ is given by \re{YM1n-1} and ${\cal L}_{\rm CS}^{(n)}$ is the CS density in $d=2n+1$ dimensional
spacetime.
  
The definition of the Chern-Simons densities is standard, if complicated, but here we are interested only in $static$
field configurations, in which case it simplifies considerably. The relevant formulas are presented in the Appendix, and
here we simply state the simplified definition of the static CS densities. With $\partial_t A_{\mu}=0$, one can show that,
up to a total divergence term (which we ignore here since we are only interested in the Euler-Lagrange equations),
the effective arbitarary $n$ CS Lagrangian reduces to the effective density
\be
\label{CSstat}
{\cal L}_{\rm CS}^{(n)}=(n+1)\,\vep^{i_1i_2i_3i_4\dots i_{2n-1}i_{2n}}\mbox{Tr}
\bigg\{ 
\,A_0\,F_{i_1i_2}\,F_{i_3i_4}\dots
F_{i_{2n-1}i_{2n}}
\bigg\}
\,.
\ee
To our knowledge, this is a new result, and its derivation is given in Appendix A.

Finally, we define the scalar valued global charges of our solutions.
The global charges of a dyon are the magnetic and the electric fluxes. The ``magnetic''
flux on $\R^{2n}$ here is the $n$-th Chern-Pontryagin charge  ${\cal C}^{(n)}$,
appearing $e.g.$ in \re{topineq} and \re{topineq1}. The definition is familiar, in
terms of the static ``magnetic'' curvature $F_{ij}$.

Next, definition of the electric flux.
As always, the appearance of a Chern-Simons term in the Lagrangian gives
rise to a nontrivial timelike, or ``electric'',
component of the gauge field. The electric flux here
\be
\label{elecflux}
q\simeq\int\,dS_i\,E_{i}\,,
\ee
($E_i=F_{i0}$) is a non-Abelian quantity.
However, this flux takes its values always in one single element of the
$SO(D+2)$ algebra, and can therefore be interpreted as the global electric charge. The reason is that at infinity the field
is radially symmetric, and subject to spherical symmetry the the solutions are restricted to the
$SO(D)\times{SO(2)}$ subalgebra of $SO(D+2)$, and, $A_0$ takes its values along the $SO(2)$ subalgebra. This fact
will become clear from the asymptotic analysis below (see \re{asympt-inf}).

An alternative definition for the electric flux can be given such that it is
expressed as a scalar valued global charge, {\it ab initio}. This is in the same
spirit
as in \cite{Lambert:1999ua}, the only difference here being that
we do not have a Higgs field, instead of which we employ the non-Abelian gauge connection $A_0$
\be
\label{electric}
q\stackrel{\rm def.}\simeq\int\,dS_i\,\mbox{Tr}\, \bigg\{ A_0\,E_{i} \bigg\} =\int\,dS_i\,\mbox{Tr}\,A_0\,D_iA_0=\frac12\,\int\,dS_i\,\pa_i\mbox{Tr}\, \bigg\{ A_0^2 \bigg\}
=\frac12\,\int\,d\Omega_{(D)}\,\frac{d}{dr}A_0^2\,,
\ee
where $d\Omega_{(D)}$ is the angular volume element in $\R^D$.
 Both definitions \re{elecflux} and \re{electric} give, up to normalisation, the same result.
 
Before proceeding to the consideration of specific models considered, we return to our statement in footnote $1$, namely that the
choice of YM sector ${\cal L}_{\rm YM}$ in \re{YMCS}, given by $p_1=p_2=\frac{D}{4}$ in \re{topineq}, is excluded in the present work.
There exist no finite energy solutions to those systems, but as seen in the $D=4$ ($n=2$) case in $5$-dimensional spacetime such solutions can be
constructed when a suitable scalar field is introduced~\cite{Collie:2008vc}.
Whether such solutions can be constructed in higher dimesnional spacetimes with $n\ge 3$
by the introduction of a scalar field is an open question.

In our concrete numerical constructions,
we will consider the following two actions, in $d=6+1$ and $d=8+1$ spacetime
dimensions, $n=3$ and $n=4$, with $D=2n$.
\bea
S_7&=& \int_{ \mathcal{M}}  d^7x \sqrt{-g} \bigg [
\frac{\tau_1}{2\cdot 2!}\,\mbox{Tr} \big\{ F_{\mu\nu}F^{\mu\nu} \big \} 
+\frac{\tau_2}{2\cdot 4!}\,\mbox{Tr} \big\{ F_{\mu\nu\rho\si}F^{\mu\nu\rho\si} \big \}   \bigg ] 
+\kappa \int_{ \mathcal{M}}  d^7 x~ {\cal L}_{\rm{CS}}^{(3)}\label{action7}\\
S_9&=& \int_{ \mathcal{M}}  d^9x \sqrt{-g} \bigg [ 
\frac{\tau_1}{2\cdot 2!}\,\mbox{Tr} \big\{ F_{\mu\nu}F^{\mu\nu} \big \} 
+\frac{\tau_2}{2\cdot 6!}\,\mbox{Tr} \big\{ F_{\mu\nu\rho\si\tau\la}F^{\mu\nu\rho\si\tau\la} \big \}  \bigg ] 
+\kappa \int_{ \mathcal{M}}  d^9 x~ {\cal L}_{\rm{CS}}^{(4)}\label{action9}
\eea
where  
$F_{\mu\nu}=\partial_\mu A_\nu-\partial_\nu A_\mu+ [A_\mu,A_\nu]$
is the gauge field strength tensor, and $\kappa$ is the CS coupling constant.

\section{The Ansatz and field equations}
In Appendix B, the static $SO(D+2)$ YM system on $\R^{D}$ ($D=2n$) is subjected to spherical symmetry.
In the present work, we use a particular truncation of this Ansatz, namely one where the functions $\vec\f$ and $\vec\chi$
(which define the magnetic and electric potentials, respectively)
are replaced by
\[
\vec\f=(w,0,0)\quad{\rm and}\quad\vec\chi=(V,0,0)\,.
\]
This is because subject to spherical symmetry, we could construct only $SO(D)\times{SO(2)}$ solutions,
and not the ones for the full $SO(D+2)$ gauge group. 
We have presented the general case in Appendix B
since we expect that introduction of a negative cosmological constant ($i.e.$ an Anti-de Sitter spacetime background) would enable the
systematic extension of the solution presented here to the full $SO(D+2)$ 
case \footnote{In \cite{Brihaye:2009cc} (for $D=4$), the presence of a negative cosmological constant
resulted in the full $SO(D+2)$ solutions being realised for the gravitating YMCS system. }.
The $SO(D)\times{SO(2)}$ truncated version of the Ansatz employed in this work is
\begin{eqnarray}
\label{YMansatz}
A= 
\frac{w(r)+1}{r} \Sigma_{ij}\frac{x^i}{r}\,dx^j
+V(r)\Sigma_{D+1,D+2}\ dt
,~~{\rm with~~} i,j=1, \dots,D\,,
\end{eqnarray}
$\Sigma_{ij}$ being the chiral representation matrices of $SO(D)$, and $\Sigma_{D+1,D+2}$ of the $SO(2)$, subalgebras
in $SO(D+2)$, defined in Appendix B. $r$ and $t$ are the radial and time coordinate
in the $(D+1)$-dimensional Minkowski space, while $x^i$ are the usual Cartesian coordinates, with $x^ix^i=r^2$.
it is convenient to use the following chiral Sigma matrices in $(D+2)$ dimensions,
\be
\label{chiralSiD}
\Si_i=-\tilde\Si_i=i\,\Ga_i\ ,\quad\Si_{D+1}=-\tilde\Si_{D+1}=i\,\Ga_{D+1}\ ,\quad\Si_{D+2}=\tilde\Si_{D+2}=\eins
\ee
in terms of the gamma matrices $\Ga_M=(\Ga_i,\Ga_m)$ in $(D+2)$ dimensions.

Subject to the spherical symmetry Ansatz \re{YMansatz} the $p$-YM ansatz density on $\R^D$ reduces to the following
one dimensional functional of $r$
\bea
\label{F2ptrunc}
\mbox{Tr} \bigg\{  F(2p)^2 \bigg\}
\sqrt{-g}
&\simeq&\frac{(D-1)!}{(D-2p+1)!}\
r^{D-1}\left(\frac{w^2-1}{r^2}\right)^{2(p-1)}\bigg\{
-(2p)\,V'^2+\nonumber\\
&&+(2p-1)\,(D-2p+1)\,\left[(2p)\,\left(\frac{w'}{r}\right)^2+(D-2p)\left(\frac{w^2-1}{r^2}\right)^2\right]\bigg\}\,,
\eea
where the power of $r^{D-1}$ in the volume element is included.

The corresponding one dimensional residual density pertaining to \re{CSstat} for arbitrary $D=2n$ can readily be calculated
\be
\label{CSstatred}
L_{(\rm CS)}^{(n)}=\ka\,(n+1)\,V\,(w^2-1)^{n-1}\,w'\,.
\ee

Substituting this Ansatz in \re{YMCS}, we find the one-dimensional reduced Lagrangian
\begin{eqnarray}
\label{Leff}
S_{eff}=&&\int dt dr ~\bigg[
r^{D-1}
\left(
-\frac{1}{2}V'^2+(D-1)       \left(\frac{w'^2}{2r^2}
+\frac{(D-2)(1-w^2)^2}{4 r^4} \right)
\right)+
\\
\nonumber
&&{~~~~~~~~}
+\frac{\tau}{2(2p-1)!}r^{D-1}\left(\frac{1-w^2}{r^2} \right)^{2(p-1)}
\left (-V'^2+(2p-1)(D-2p+1)\left(
\frac{w'^2}{r^2}
+\frac{(D-2p)(1-w^2)^2}{2pr^4}
\right)
\right)
\\
\nonumber 
&&{~~~~~~~~}-\kappa D(D+2)V(w^2-1)^{\frac{1}{2}(D-2)}w'
\bigg],
\end{eqnarray}
where a prime denotes a derivative with respect to $r$.
Here we have scaled out the factor $\tau_1$ and denoted $\tau=\tau_{2}\frac{(d-2)!}{(d-2p)!}$;
also to simplify the relations we have denoted $p_2=p$ henceforth.

The corresponding energy of the solutions is given by the integral
\begin{eqnarray}
\label{Energy}
&&E=V_{D-1}\int_0^\infty dr ~\bigg[
r^{D-1}
\left(
 \frac{1}{2}V'^2+(D-1)       \left(\frac{w'^2}{2r^2}
+\frac{(D-2)(1-w^2)^2}{4 r^4} \right)
\right)
\\
\nonumber
&&{~~~~~~~~}
+\frac{\tau}{2(2p-1)!}r^{D-1}\left(\frac{1-w^2}{r^2} \right)^{2(p-1)}
\left (V'^2+(2p-1)(D-2p+1)
\left(
\frac{w'^2}{r^2}
+\frac{(D-2p)(1-w^2)^2}{2pr^4}
\right)
\right)
\bigg].
\end{eqnarray}
The resulting (ordinary differential) equations are 
\bea
\label{w''eq}
&&
w''
\left(1+\tau \frac{(D-2p+1)}{(D-1)}\frac{1}{(2p-2)!} 
(\frac{1-w^2}{r^2})^{2(p-1)}
\right)
+\frac{(D-3)w'}{r}
-\frac{(D-2)w(w^2-1)}{r^2}-
\\
\nonumber
&&
-\frac{D(D+2)\kappa}{r^{D-3}(D-1)}\frac{dF(w)}{dw}V'
+\tau\frac{1}{(2p-1)!}\frac{2(p-1)}{D-1}(\frac{w^2-1}{r^2})^{2p-3}
\bigg(
wV'^2+
\\
\nonumber
&&
+
\frac{(2p-1)(D-2p+1)}{2(p-1)r^4}(rw'((D-4p+1)(w^2-1)+2(p-1)rww')
-(D-2p)w(1-w^2)^2)
\bigg)
=0,
\eea
and
\be
\label{V'eq0}
\left[ r^{D-1} V'\left(1+\frac{\tau}{(2p-1)!}(\frac{1-w^2}{r^2})^{2(p-1)}\right)
-\kappa D(D+2)F(w)
\right ]' =0\,,
\ee
the last one being the Gauss Law equation.
The function $F(w)$ in \re{w''eq}-\re{V'eq0} has the following general expression in terms of the hypergeometric function
 $_2F_1$:
\begin{eqnarray}
F(w)=-(-1)^{\frac{D}{2}}~_2F_1\left(\frac{1}{2},\frac{2-D}{2},\frac{3}{2};w^2\right)w.
\end{eqnarray} 

As seen from the  relation (\ref{V'eq0}), an important generic feature of YMCS models within the $SO(D)\times SO(2)$
truncation is the existence of a first integral  for the electric potential $V(r)$,
\be
\label{V'eq}
V'+\frac{1}{r^{D-1}}\frac{P-\kappa D(D+2)}{1+\frac{\tau}{(2p-1)!}(\frac{w^2-1}{r^2})^{2(p-1)}}F(w)=0,
\ee
with $P$ an integration constant, which, as we shall see, for regular solutions is fixed by $\kappa$.
 
One can note that equations \re{w''eq} and \re{V'eq}
are invariant under the scaling
\begin{eqnarray}
\label{ss1}
 r\to \lambda r, ~w\to w,~~V\to V/\lambda,~~\tau \to \lambda^{4(p-1)} \tau,~~
{\rm and}~~\kappa \to \lambda^{d-4} \kappa, 
\end{eqnarray}
with $\lambda$ an arbitrary positive parameter. 
Then, without any loss of generality. one can set in this way   $\tau$ or $\kappa$
to take an arbitrary nonzero value.

Another symmetry of the equations of the model (\ref{w''eq}), (\ref{V'eq0}) consists in simultaneously
changing the sign of the CS coupling constant together with the electric or magnetic potential
\begin{eqnarray}
\label{ss3}
\kappa \to -\kappa,~~V\to - V,~~~~{\rm or}~~~\kappa \to -\kappa,~~w\to - w.
\end{eqnarray} 
In what follows, we shall use this symmetry to study solutions with a positive 
$\kappa$ only.

\section{The solutions}

 
 We start by presenting the expansion of the solutions at near the origin  $r=0$.
 The regularity of the gauge field implies $w\to \pm 1$ there.
 Then it follows that the  parameter $P$ in the first integral (\ref{V'eq}) is
fixed by the value of the CS coupling constant. Technically,
this results from the fact that the term $P-\kappa D(D+2)F(\pm 1)$ should vanish as
$r\to 0$.
Restricting without any loss of generality to 
$w(0)=1$, one finds
\begin{eqnarray}
\label{P}
P=\kappa D(D+2)F(1).
 \end{eqnarray}
 Then one finds  the following expansionsnear the origin:
\begin{eqnarray}
\label{origin}
&&w(r)= 1-br^2+O(r^4), 
~~V(r)= \kappa\frac{(-2)^{\frac{D}{2}}b^{\frac{D+4}{2}}\Gamma(2p+1)}{(2b)^{2p}\tau+2b^2 \Gamma(2p+1)} r^2+O(r^4),~~ 
 \end{eqnarray}
 The only free parameters here is $b=-\frac{1}{2}w''(0)$.
The coefficients of all higher order terms in the $r\to 0$ expansion are fixed by $b$.

 The solutions have the following expansion\footnote{Note that although $w(\infty)=1$ is also allowed, we could not find such $multinode$
 solutions.}  as $r\to \infty$:
\begin{eqnarray}
\label{asympt-inf}
w(r)=-1-\frac{J}{r^{D-2}}+O(1/r^D),~~
V(r)=V_0-\frac{Q}{r^{D-2}}+O(1/r^D).
\end{eqnarray}
In the above relations,  $J,~V_0$ are parameters given by numerics which fix all higher order terms,
while $Q$ is a constant fixing the electric charge of the solutions,
\begin{eqnarray}
Q=\kappa (-1)^{\frac{D}{2}}\sqrt{\pi}\frac{D(D+2)}{D-2}\frac{\Gamma(\frac{D}{2})}{\Gamma(\frac{D+1}{2})}.
 \end{eqnarray}
%


The solutions interpolating  between
the asymptotics (\ref{origin}), (\ref{asympt-inf})  were constructed by using a standard Runge-Kutta
ordinary  differential equation solver. In this approach we 
evaluate the initial conditions at $r= 10^{-5}$, for global tolerance $10^{-12}$,
adjusting  for shooting parameters and integrating  towards  $r\to\infty$. 
The input parameters are $\tau$ and $\kappa$ (note that only the ratio of these 
two constant is relevant).

Our numerical analysis gives evidence for the existence of nontrivial solutions of the
eqs. (\ref{w''eq}), (\ref{V'eq0}) for $D=6,8$, these cases being studied in a systematic way. 
However, solutions with similar properties should exist for all higher values of $D=2n$.
The profile of a typical  solution for the case $D=6,~p=2$ is given in Figure 1.
For all solutions
 we have constructed, the  gauge function $w(r)$
 has a monotonic behaviour, with a single node for an intermediate value of $r$.
 A monotonic behaviour has been also noticed for the electric potential $V(r)$.
 For moderate values of the ratio $\tau_2/\kappa$, the energy density is mainly concentrated
 in a shall, its maximum corresponding to the position of the of the node of the 
 magnetic gauge function.  
  
\setlength{\unitlength}{1cm}
\begin{picture}(8,6)
\put(-0.5,0.0){\epsfig{file=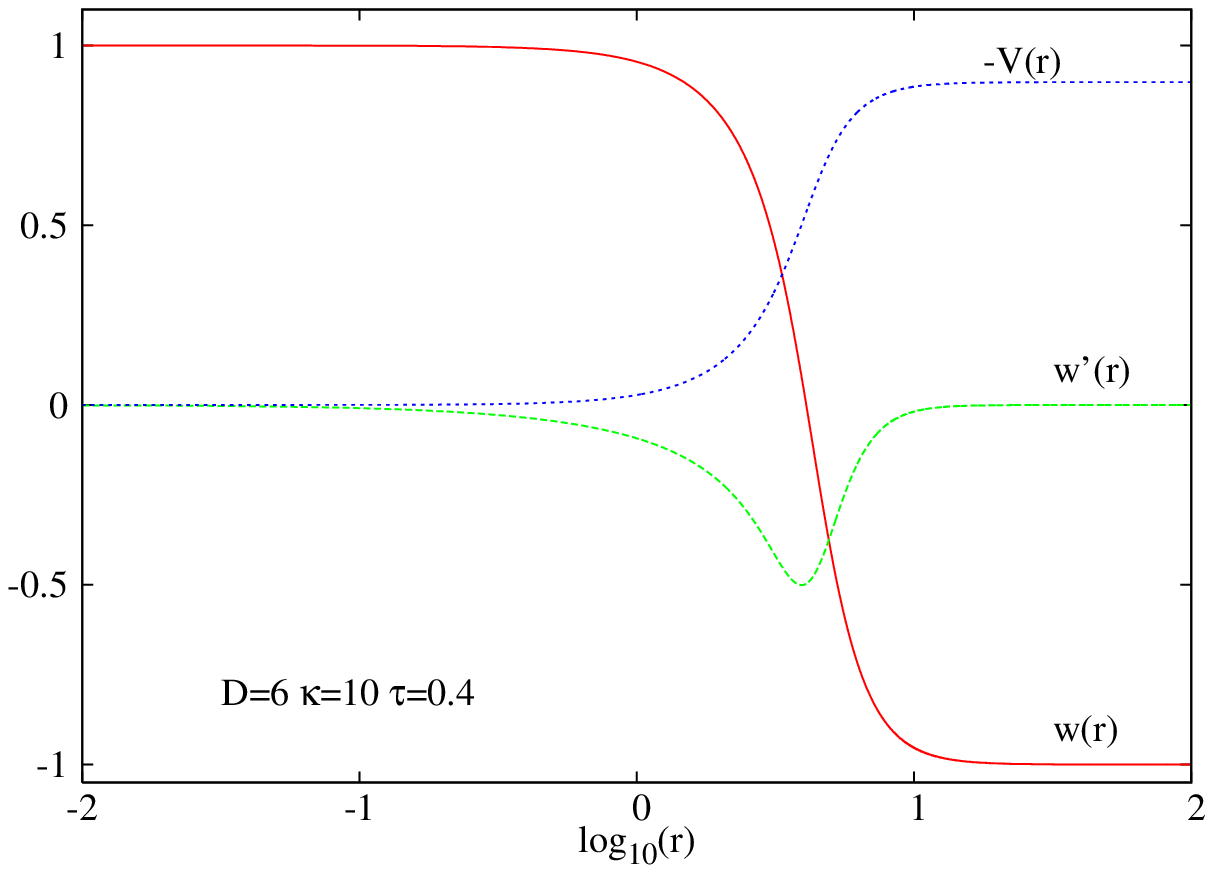,width=8cm}}
\put(8,0.0){\epsfig{file=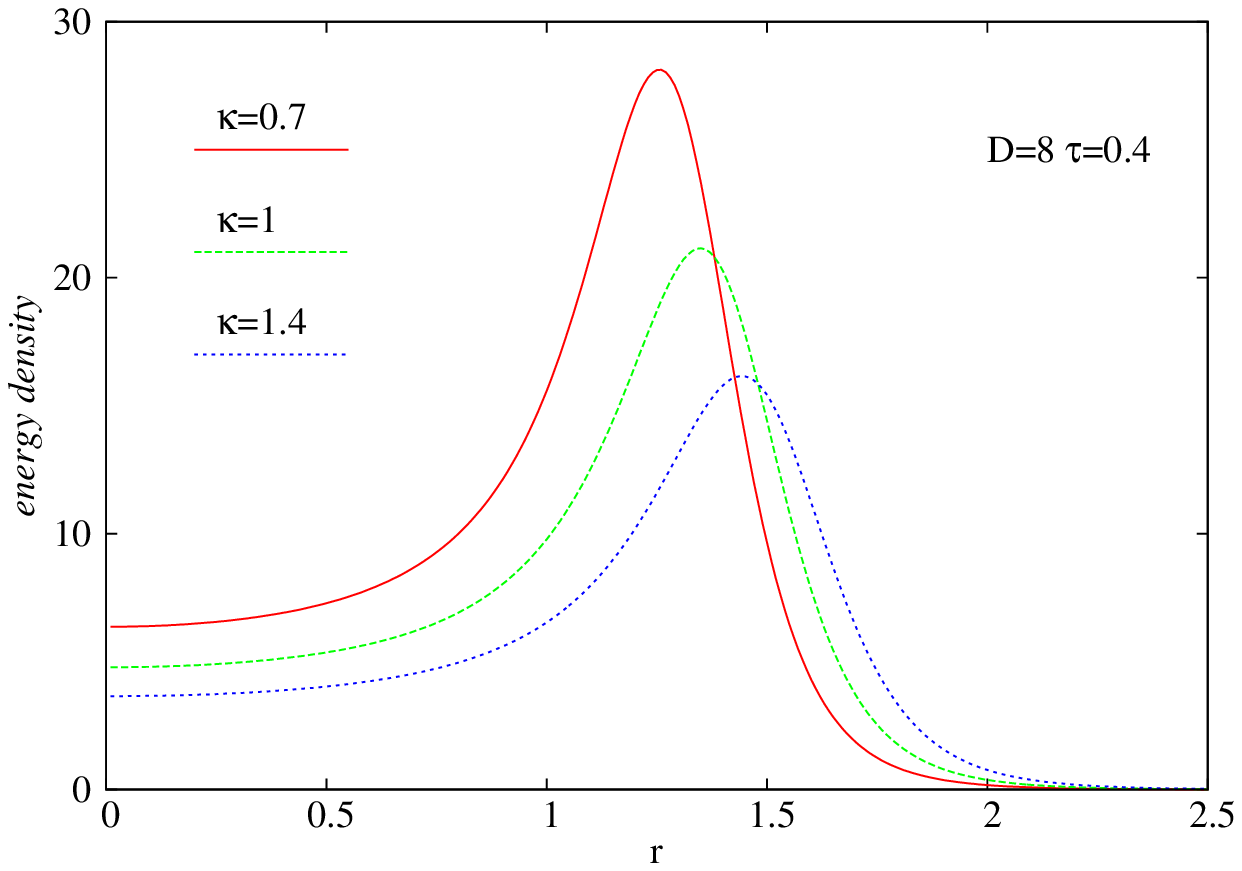,width=8cm}}
\end{picture}
\\
\\
{\small {\bf Figure 1.} {\it Left:} The profiles of a $D=6$ solution are shown as a function of the radial coordinate.
{\it Right:}
The energy density is shown for $D=8$ solutions with different values of $\kappa$ and the same $\tau$.
   }
\vspace{0.5cm}

\setlength{\unitlength}{1cm}
\begin{picture}(8,6)
\put(-0.5,0.0){\epsfig{file=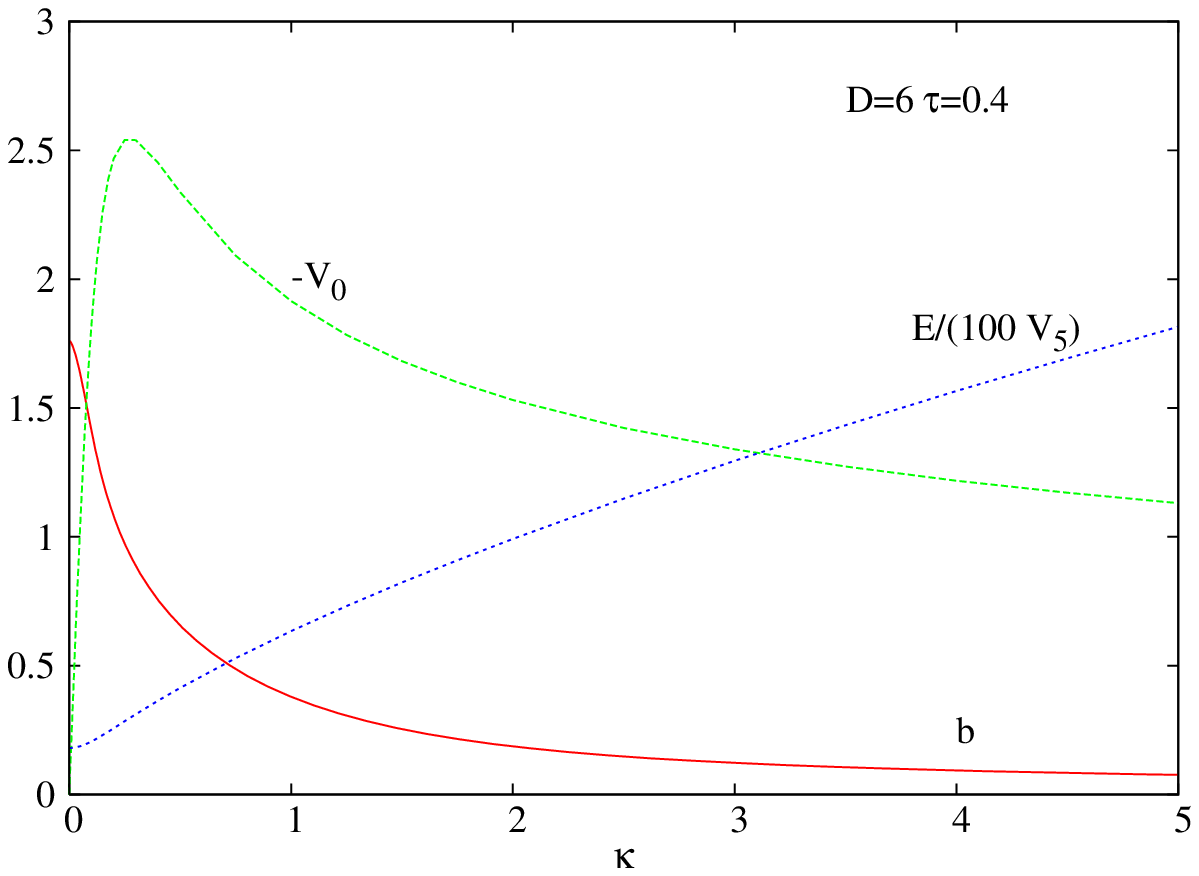,width=8cm}}
\put(8,0.0){\epsfig{file=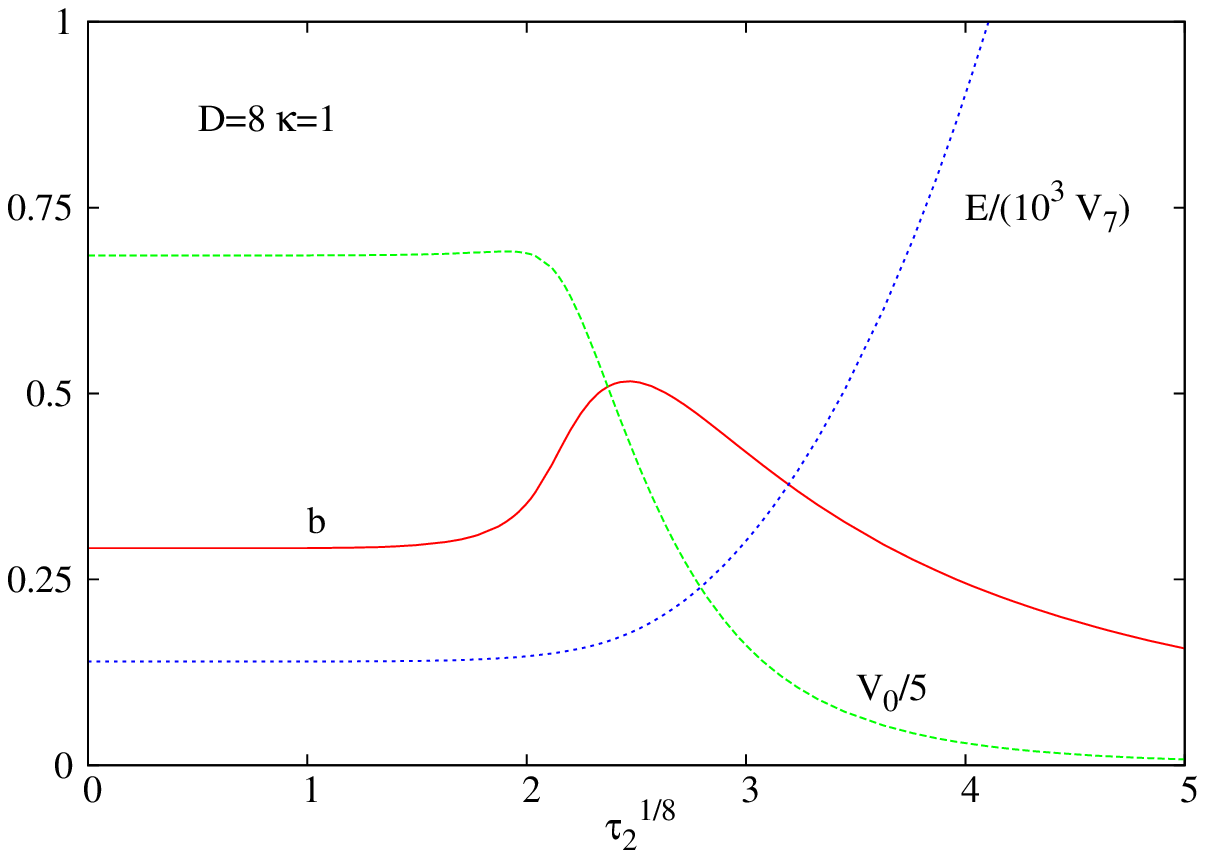,width=8cm}}
\end{picture}
\\
\\
{\small {\bf Figure 2.}  The total energy $E$, the shooting parameter $b=-\frac{1}{2}w''(0)$, and the asymptotic value  $V_0$
of the electric potential are shown as a function of the parameter $\kappa$ ($left$) and as a function of the parameter $\tau_2$
($right$) for $D=6,8$ solutions, respectively.

   }
\vspace{0.5cm}

The dependence of solutions on the parameters $\kappa,\tau_2$ is shown in Figure 2.
Starting with the dependence on the parameter $\kappa$, we notice that nontrivial solutions 
with finite mass are also found in the absence of a CS term, $\kappa=0$.
However, these configurations have a vanishing electric potential
and correspond to the instantons in \cite{Burzlaff:1993kf}.
The total mass of solutions increases almost linearly with $\kappa$, while 
the shooting parameter $b=-w''(0)/2$ decreases.
Also, as one can see in Figure 2 (right), one can find finite energy solutions of the YMCS
equations even in the absence of the ${\cal L}_{\rm YM}^{(n-1)}$ term in (\ref{YM1n-1}),
$i.e.$ for $\tau_2=0$.

\section{Summary and discussion}
We have formulated the construction of dyon solutions to a family of
Yang-Mills--Chern-Simons (YMCS) systems in $2n+1$ ($n\ge 3$)
dimensional flat spacetime. These are
static spherically symmetric field configurations carrying both magnetic and electric
flux. The components of the non-Abelian gauge connection $A_{\mu}=(A_i,A_0)$ are
defined to be in the algebra of $SO(D+2)$, but subject to spherically symmetry
the solutions turn out to take their values in the
subalgebra $SO(D)\times SO(2)$.

Just as for the other known dyonic solutions in four~\cite{Julia:1975ff} and
five~\cite{Lambert:1999ua}
dimensional spacetimes, the ``magnetic'' flux is a topological charge.
In four dimensions the topological charge of the Julia-Zee (JZ) dyon
is the monopole charge on $\R^3$, which is a descendant~\cite{Tchrakian:2010ar} of
the second Chern-Pontryagin (CP) charge. In five dimensions the topological charge of the
Lambert-Tong (LT) dyonic instanton, it is the second CP charge itself on $\R^4$.
In our case here, the topological charge is the $n$-th CP charge on $\R^n$, $n\ge 3$.
In the above, we have referred to the topological charges as ``magnetic'' charges,
in analogy with \cite{Julia:1975ff}, and also because these are defined exclusively
in terms of the static spacelike (or ``magnetic'') components of the connection $A_i$, with
the timelike (or ``electric'') components $A_0$ entering the definition of the ``electric''
flux.

In this context the JZ dyon can be described as a $monopolic$
dyon~\footnote{Configurations carrying non-Abelian ``electric'' fields in all dimensions can
be readily constructed by {\it riding on the back} of monopoles on $\R^D$. But the existence
of a scalar valued global electric flux relies on the definition of a '~Hooft electromagnetic,
something that is problematic in flat spacetime dimensions higher than $4$.} while the LT dyon
and our solutions can be described as a $instantonic$ dyons. The salient difference between
the two types of topological charge is that the former decay like
Dirac-Yang~\cite{Tchrakian:2008zz}
monopoles and hence in all but $3+1$ dimensions the energy
integral of the usual (quadratic) Yang-Mills (YM) term diverges,
while the latter decay (like instantons) faster and the ``energy''
integral of the usual YM term converges. This can be seen as a desirable 'physical' feature.

Concerning the detailed choice of the models proposed, the gauge field system supporting the
topological (``magnetic'') charge is directly specified like in the known dyonic models.
In the case of the JZ dyon, this is the usual
$SU(2)$ Georgi-Glashow model~\footnote{Though
in that case one can employ instead any one of the $excited$
Georgi-Glashow models, or any superposition thereof, which descend from higher than $4$
dimensions~\cite{Tchrakian:2010ar}.}. In the case of the LT dyon, this is the usual (quadratic)
$SO_{\pm}(4)=SU_{L/R}(2)$ Yang-Mills model. In our case, like LT, the sector supporting the
topological charge is described by the non-Abelian gauge field only, and the natural system
is then the superposition of members of the Yang-Mills hierarchy \re{pYMsuperposn}. Now the
number of terms in \re{pYMsuperposn} grow with increasing $D$ (or $n$) and it may be
reasonable to retain the minimum number of these necessary for the Derrick scaling requirement
to be satisfied. For this, one can select any pair of terms, both with even $p$, for which
the topological inequality \re{topineq1} can be satisfied. Of such pairs, it may be reasonable
to privilege the pair $p_1=1\,,\, p_2=n-1$, if only for ``physical'' reasons favouring
the presence of the usual (quadratic) YM term. This is not necessary, but it is what we have
done and as it happens in the two examples in $7$ and $9$ dimensions
studied quantitatively, these are the
unique possibilities.

$Alternatively$, one might have opted to retain only the $p=\frac{D}{4}$ term in
\re{pYMsuperposn}, $i.e.$ with $\tau_1=0$ in (\ref{YM1n-1}), which on its own satisfies the Derrick
scaling requirement. (See footnote $1$.) After the introduction of the Chern-Simons (CS) term
(and of $A_0$) however, such solutions were not found. This is because the appearance of the CS term
in the action results in the effective presence of a higher order term in the static energy
density functional, as explained in Section {\bf 1}. This effective higher order CS term, which scales as $L^{-1}$, then destroys the previous
Derrick scaling balance.
This is why we have proposed models in $d=2n+1$ for $n\ge 3$ only, since in $d=5$ the only choice for the (``magnetic'') YM sector is the
$p=1$ YM term. How this obstacle is circumvented in the case of LT dyonic instanton~\cite{Lambert:1999ua} (and all higher analogues in
$d=4p+1$ dimensional spacetime) can be seen in Appendix {\bf C}.

As a final comment, we note that the solutions discussed here are exclusively spherically symmetric.
One would expect that there exist dyons of these systems subject to less stringent symmetries, though
technically the numerical construction of these putative dyons would be a formidable task.

\bigskip
\bigskip
\noindent
{\bf\large Acknowledgements}

\noindent
We thank Neil Lambert, David Tong and Yisong Yang for discussions on dyons of non-Abelian Yang-Mills Chern-Simons systems.
This work is carried out in the framework of Science Foundation Ireland (SFI) project
RFP07-330PHY.  
E.R. gratefully acknowledges support by the DFG.

\appendix
\section{Static limit of the Chern-Simons density in $2n+1$ dimensions}
\setcounter{equation}{0}
\renewcommand{\theequation}{A.\arabic{equation}}

To illustrate our statement, namely the result \re{CSstatred},
we carry out the relevant computations explicitly in $3,\,5$ and $7$ dimensional spacetimes.
The CS densities $\Omega_{\rm CS}^{(2n+1)}$ in question, for $n=1,2$ and $3$, are
\bea
\nonumber
\Omega_{\rm CS}^{(3)}&=&\vep_{\la\mu\nu}\mbox{Tr}\, \bigg\{
A_{\la}\left[F_{\mu\nu}-\frac23A_{\mu}A_{\nu}\right]\bigg\},
\label{CS3}
\\
\Omega_{\rm CS}^{(5)}&=&\vep_{\la\mu\nu\rho\si}\mbox{Tr}\,
\bigg\{
A_{\la}\left[F_{\mu\nu}F_{\rho\si}-F_{\mu\nu}A_{\rho}A_{\si}+
\frac25A_{\mu}A_{\nu}A_{\rho}A_{\si}\right]\bigg\}
,
\label{CS5}
\\
\Omega_{\rm CS}^{(7)}&=&\vep_{\la\mu\nu\rho\si\tau\ka}
\mbox{Tr}\,
\bigg\{ A_{\la}\bigg[F_{\mu\nu}F_{\rho\si}F_{\tau\ka}
-\frac45F_{\mu\nu}F_{\rho\si}A_{\tau}A_{\ka}-\frac25
F_{\mu\nu}A_{\rho}F_{\si\tau}A_{\ka}
\nonumber
\\
\nonumber
&&\qquad\qquad\qquad\qquad\qquad\qquad
+\frac45F_{\mu\nu}A_{\rho}A_{\si}A_{\tau}A_{\ka}-\frac{8}{35}
A_{\mu}A_{\nu}A_{\rho}A_{\si}A_{\tau}A_{\ka}\bigg]\,
\bigg\},\label{CS7}
\eea
which are by construction $gauge\ variant$. Their variations however are by contrast $gauge\ covariant$
and are easily expressed in
general $2n+1$ case as
\be
\label{E-L_CS}
\delta_{A_{\la}}\Omega_{\rm CS}^{(2n+1)}=(n+1)
\,\vep^{\la\mu_{1}\mu_{2}\mu_{3}\mu_{4}\dots\mu_{2n-1}\mu_{2n}}\,F_{\mu_1\mu_2}\,F_{\mu_3\mu_4}\dots\,F_{\mu_{2n-1}\mu_{2n}}\,.
\ee
The statement here is the following: {\it for static fields, the CS density $\Omega_{\rm CS}^{(2n+1)}$ on $d=2n+1$ dimensional
spacetime reduces to \re{CSstatred}, namely to}
\be
\label{staticCS2n+1}
\Omega_{\rm CS}^{(2n+1)}=\bnabla\cdot\bOmega^{(2n+1)}+(n+1)\mbox{Tr}\,\bigg\{ A_0\,F\wedge F\wedge\dots\wedge F\,  \bigg\},
\quad n\ {\rm times}\,.
\ee
To illustrate this, we give the explicit total divergence expressions for $n=1,2$ and $3$,
\re{staticCS2n+1} are,
\bea
\nonumber
\Omega_{\rm CS}^{(3)}&=&2\,\vep_{ij}\mbox{Tr}\, \bigg\{ A_0\,F_{ij} \bigg\}+\bnabla\cdot\bOmega^{(3)}\label{staticCS3}
\\
\Omega_{\rm CS}^{(5)}&=&3\,\vep_{ijkl}\mbox{Tr}\, \bigg\{ A_0\,F_{ij}\,F_{kl} \bigg\}+\bnabla\cdot\bOmega^{(5)}
\label{staticCS5}
\\
\nonumber
\Omega_{\rm CS}^{(7)}&=&4\,\vep_{ijklmn}\mbox{Tr}\, \bigg\{A_0\,F_{ij}\,F_{kl}\,F_{mn} \bigg\}
+\bnabla\cdot\bOmega^{(7)}\,,\label{staticCS7}
\eea
with $\bOmega^{(2n+1)}\equiv\Omega_i^{(2n+1)}$ given by
\bea
\nonumber
\Omega_i^{(3)}&=&-2\,\vep_{ij}\mbox{Tr}\, \bigg\{A_0\,A_j\bigg\},
\label{omega3}
\\
\Omega_i^{(5)}&=&-2\,\vep_{ijkl}\mbox{Tr} \bigg\{A_0\left[(A_j\,F_{kl}+F_{kl}\,A_j)-A_j\,A_k\,A_l\right] \bigg\},
\label{omega5}
\\
\nonumber
\Omega_i^{(7)}&=&-2\,\vep_{ijkl}\mbox{Tr}\, \bigg\{ A_0\bigg[(F_{ij}\,F_{kl}\,A_n+F_{ij}\,A_n\,F_{kl}+A_n\,F_{ij}\,F_{kl})
\\
\nonumber
&&\qquad\qquad\qquad-\frac45(A_k\,A_l\,A_n\,F_{ij}+F_{ij}\,A_k\,A_l\,A_n)\label{omega7}\\
&&\qquad\qquad\qquad-\frac25(A_k\,A_l\,F_{ij}\,A_n+A_n\,F_{ij}\,A_k\,A_l)
+\frac45 A_i\,A_j\,A_k\,A_l\,A_n\bigg]
\bigg\}
\,.\nonumber
\eea

\section{Spherically symmetric Ansatz for $SO(D+2)$ YM on $\R^D$}
\setcounter{equation}{0}
\renewcommand{\theequation}{B.\arabic{equation}}

The static spherical symmetric Ansatz for the $SO(6)$ Yang-Mills (YM) system on $\R^6$, used in
\cite{Brihaye:2009cc}, is extended to the $SO(D+2)$ YM on $\R^D$,
$i.e.$ in $D+1$ dimensional spacetime with $D=2n$. Since $D$ is even,
we can take the YM connection to take its values in one or other
chiral representation of $SO_{\pm}(D+2)$, such that our spherically
symmetric Ansatz is expressed in terms of the representation matrices
\be
\label{sigmap}
\Sigma_{\al\beta}^{(\pm)}=-\frac{1}{4}\left(\frac{1\pm\Gamma_{2n+3}}{2}\right)
[\Gamma_{\al} ,\Gamma_{\beta}]\quad,\quad \al,\beta=1,2,...,2n+2\ ,
\ee
$\Gamma_{\al}=(\Gamma_{i},\Gamma_{M})$, with the index $M=(2n+1,2n+2)$, being
the gamma matrices in $2n+2$ dimensions and $\Gamma_{2n+3}$,
the corresponding chiral matrix.

Our spherically symmetric Ansatz for the YM connection $A_{\mu}=(A_0,A_i)$ is
\bea
A_0&=&-(\vep\chi)^M\,\hat x_j\,\Sigma_{jM}^{(\pm)}-
\chi^{2n+3}\,\Sigma_{2n+1,2n+2}^{(\pm)}\label{a0p}\\
A_i&=&\left(\frac{\f^{2n+3}+1}{r}\right)\Sigma_{ij}^{(\pm)}\hat x_j+
\left[\left(\frac{\f^M}{r}\right)\left(\delta_{ij}-\hat x_i\hat x_j\right)+
(\vep A_r)^M\,\hat x_i\hat x_j\right]\Sigma_{jM}^{(\pm)}+\nonumber\\
&&\qquad\qquad\qquad\qquad\qquad\qquad\qquad\qquad +A_r^{2n+3}\,
\hat x_i\,\Sigma_{2n+1,2n+2}^{(\pm)}\label{aip}
\eea
in which the summed over indices $M,N=2n+1,2n+2$ run over two values such that
we can label the functions $(\f^M,\f^{2n+3})\equiv\vec\f$,
$(\chi^M,\chi^{2n+3})\equiv\vec\chi$ and $(A_r^M,A_r^{2n+3})\equiv\vec A_r$
like three isotriplets $\vec\f$, $\vec\chi$ and $\vec A_r$, all depending on
the $2n$ dimensional spacelike radial variable $r$. $\vep$ is the two
dimensional Levi-Civita symbol.

The parametrisation used in the Ansatz \re{a0p}-\re{aip} results in a gauge
covariant expression for the YM curvature $F_{\mu\nu}=(F_{ij},F_{i0})$
\bea
F_{ij}&=&\frac{1}{r^2}\left(|\vec\f|^2-1\right)\Sigma_{ij}^{(\pm)}+
\frac1r\left[D_r\f^{2n+3}+\frac1r\left(|\vec\f|^2-1\right)\right]
\hat x_{[i}\Sigma_{j]k}^{(\pm)}\hat x_{k}+
\frac1rD_r\f^M\hat x_{[i}\Sigma_{j]M}^{(\pm)}\label{fijp}\\
F_{i0}&=&-\frac1r\,\f^M(\vep\chi)^M\,\Sigma_{ij}^{(\pm)}\hat x_j+\frac1r\,
\left[\f^{2n+3}(\vep\chi)^M-\chi^{2n+3}(\vep\f)^M\right]\Sigma_{iM}^{(\pm)}
\nonumber\\
&-&\left\{(\vep D_r\chi)^M+\frac1r\,
\left[\f^{2n+3}(\vep\chi)^M-\chi^{2n+3}(\vep\f)^M\right]\right\}
\hat x_i\hat x_j\Sigma_{jM}^{(\pm)}
-D_r\chi^{2n+3}\,\hat x_i\,\Sigma_{2n+1,2n+2}^{(\pm)}\label{fi0p}
\eea
in which we have used the notation
\be
\label{covp}
D_r\f^a=\pa_r\f^a+\vep^{abc}\,A_r^b\,\f^c\quad,\quad
D_r\chi^a=\pa_r\chi^a+\vep^{abc}\,A_r^b\,\chi^c
\ee
as the $SO(3)$ covariant derivatives of the two triplets
$\vec\f\equiv\f^a=(\vec\f^M,\f^{2n+3})$,
$\vec\chi\equiv\chi^a=(\vec\chi^M,\chi^{2n+3})$, with respect to the
$SO(3)$ gauge connection $\vec A_r\equiv A_r^a$. This gauge connection (in one dimension, $r$) is
of course trivial and hence does not appear in the field equations, but it serves to give the
constraint equation.

\section{Dyonic instantons in $d=4p+1$ dimensional spacetimes}
\setcounter{equation}{0}
\renewcommand{\theequation}{C.\arabic{equation}}
The dyonic instanton is a (static) solution to the Yang-Mills-Higgs model in $4+1$ dimensions. It shares with the Julia-Zee (JZ)
dyon, the presence of a Higgs term in the action, which is to be identified with the electric non-Abelian connection $A_0$
leading to the definition of the electrix flux.
This is in direct contrast with the models proposed here, which feature no Higgs field and $A_0$
appears through the Chern-Simons terms in the action.

In contrast to the JZ dyon however, static energy density functional of the Lambert-Tong (LT) dyonic instanton does not depend on
the Higgs field and consists solely of the usual Yang-Mills action density. More specifically, the YM field is the (anti-)self-dual
instanton on $\R^4$. As such the Higgs field plays an auxiliary role in the description of the LT dyonic instanton, and is needed
only for the definition of the global electric flux. This makes it possible to generalise the construction of dyonic instantons to
$4p+1$ dimensional spacetimes, albeit in a rather limited context. This is our aim in the present Appendix.

We limit our considerations here to the recovery of the $4+1$ dimensional LT dyonic instanton, and proceed to its
$8+1$ dimensional version. The $4p+1$ dimensional case then follows systematically. The Lagrangian ${\cal L}_{D+1}$ for $D=4$ is
\be
\label{L4}
{\cal L}_{4+1}=-\frac{1}{2\cdot 2!}\mbox{Tr} \big\{F_{ij}F^{ij} \big\}+\frac12\mbox{Tr} \big\{D_i\F D^i\F \big\}\,,
\ee
yielding the following gauge field equations
\bea
D_{\mu}F_{\mu 0}-[\F,D_0\F]&=&0\label{gauss4}\\
D_{\mu}F_{\mu\nu}-D_0F_{0\nu}-[\F,D_{\nu}\F]&=&0\label{ampf4}\,,
\eea
and the Higgs field equation
\be
\label{higgs4}
D_0^2\F-D_{\mu}^2\F=0
\ee
having used the Minkowskian metric $\eta_{ij}={\rm diag}(+,-,-,-,-)$, with $\eta_{00}=1$ and $\eta_{\mu\nu}=-\delta_{\mu\nu}$,
$i=0,\mu$.

Eqn. \re{gauss4} is the Gauss Law equation, and \re{ampf4}
the Ampere equations. The Hamiltonian is
\bea
{\cal H}_{4+1}&=&\frac14\mbox{Tr}\left\{\frac12 F_{\mu\nu}^2+F_{0\nu}^2+D_{\mu}\F^2+D_{0}\F^2\right \}
\label{H40}
\\
&=&\frac14\mbox{Tr}\left \{\frac12F_{\mu\nu}^2+(F_{0\mu}-D_{\mu}\F)^2+D_0\F^2
+2D_{\mu}\F F_{0\mu}\right \}
\label{H41}\,.
\eea
Following \cite{Lambert:1999ua}, we have expressed the second line of \re{H41} such, that after substituting
the Gauss Law equation \re{gauss4} and discarding the surface term arising from
\[
\pa_{\mu}\mbox{Tr} \big\{\F F_{0\mu} \big\}\,,
\]
\re{H41} takes the form
\be
\label{H4}
{\cal H}_{4+1}=\frac14\mbox{Tr}\left \{\frac12F_{\mu\nu}^2+(F_{0\mu}-D_{\mu}\F)^2+D_0\F^2\right \}\,.
\ee
We now restrict our attention to static solutions, and, choose
\be
\label{A04}
A_0=\F \implies\ \ D_0\F=0\ \ ,\ \ F_{0\mu}=D_{\mu}\F\,,
\ee
as a result of which the static limit of the Hamiltonian \re{H4} just reduces to the action density of the
$1$-YM system, $i.e.$ only the first term in \re{H4},
whose action is minimised absolutely by the $1$-BPST solution~\cite{Belavin:1975fg}.

Using \re{A04}, one can readily verify that the Ampere Law equation \re{ampf4} reduces to
\[
D_{\mu}F_{\mu\nu}=0
\]
which is solved by the selfdual BPST fields, and the Gauss
Law equation \re{gauss4} and Higgs equation \re{higgs4} become identical, reducing to
\be
\label{higgs41}
D_{\mu}D_{\mu}\F=0\,.
\ee
There remains to find a regular solution to \re{higgs41}, which we delay till after treating the $8+1$ dimensional case, next.

The Lagrangian ${\cal L}_{D+1}$ in the $D=8$ case is
\be
\label{L8}
{\cal L}_{8+1}=-\frac{1}{2\cdot 4!}\mbox{Tr} \big\{ F_{ijkl}F^{ijkl} \big\}+\frac{1}{2\cdot 3!}\mbox{Tr} \big\{F_{ijk}F^{ijk} \big\}
\ee
in which we have used the notation
\bea
F_{ijkl}&=&\{F_{i[j},F_{kl]}\}\equiv\{F_{ij},F_{kl}\}+\{F_{ik},F_{lj}\}+\{F_{il},F_{jk}\}\label{noteF4}\\
F_{ijk}&=&\{F_{[ij},D_{k]}\F\}\equiv\{F_{ij},D_{k}\F\}+\{F_{jk},D_{i}\F\}+\{F_{ki},D_{j}\F\}\label{noteF4}\,.
\eea
The choice of the Higgs kinetic term $\mbox{Tr}\{F_{[ij},D_{k]}\F\}^2$ here, instead of the usual one $\mbox{Tr}D_{i}\F^2$, is made so that
the static Euler-Lagrange equation for $A_0$ becomes identical to the corresponding Higgs equation after the identification $A_0=\F$ is
made.

Using the same Minkowskian metric as above (but in $8+1$ dimensions now) we have the following Yang--Mills equations
\bea
D_{\mu}\{F_{\rho\si},F_{0\mu\rho\si}\}+2D_{\mu}\{D_{\rho}\F,F_{0\mu\rho}\}+
[\F,\{F_{\rho\si},F_{0\rho\si}\}]&&=0\label{gauss8}\\
\nonumber\\
D_{\mu}\{F_{\rho\si},F_{\mu\nu\rho\si}\}+2D_{\mu}\{D_{\rho}\F,F_{\mu\nu\rho}\}
-2D_{\mu}\{F_{0\rho},F_{0\mu\nu\rho}\}-D_{\mu}\{D_{0}\F,F_{0\mu\nu}\}&&\nonumber\\
-D_{0}\{F_{\rho\si},F_{0\nu\rho\si}\}-2D_{0}\{D_{\rho}\F,F_{0\nu\rho}\}&&\nonumber\\
-[\F,\{F_{\rho\si},F_{\nu\rho\si}\}]-2[\F,\{F_{0\rho},F_{0\nu\rho}\}]
\ &&=0\label{ampf8}
\eea
and the Higgs equation
\be
\label{higgs8}
D_{\mu}\{F_{\rho\si},F_{\mu\rho\si}\}+2D_{\mu}\{F_{0\nu},F_{0\mu\nu}\}-D_{0}\{F_{\mu\nu},F_{0\mu\nu}\}=0\,.
\ee
Eqn. \re{gauss8} is the Gauss Law equation and \re{ampf8} the Ampere equations.

The Hamiltonian of \re{L8} is
\bea
{\cal H}_{8+1}&=&\frac{1}{4!}\mbox{Tr}\left \{\frac14
F_{\mu\nu\rho\si}^2+F_{0\mu\nu\rho}^2+F_{\mu\nu\rho}^2+2F_{0\mu\nu}^2
\right \}\label{H80}\\
&=&\frac{1}{4!}\mbox{Tr}\left \{\frac14F_{\mu\nu\rho\si}^2
+(F_{0\mu\nu\rho}-F_{\mu\nu\rho})^2+2F_{0\mu\nu}^2
+2F_{0\mu\nu\rho}F_{\mu\nu\rho}\right \}\label{H81}
\eea
where again, as in the $D=4$ case above, we have expressed the second line of \re{H81} such, that after substituting
the Gauss Law equation \re{gauss8} and discarding the surface term arising from
\[
\pa_{\rho}\mbox{Tr}\F\{F_{\mu\nu},F_{\rho\mu\nu}\}
\]
\re{H81} takes the form
\be
\label{H8}
{\cal H}_{8+1}=\frac{1}{4!}\mbox{Tr}\left[\frac14F_{\mu\nu\rho\si}^2
+(F_{0\mu\nu\rho}-F_{\mu\nu\rho})^2+2F_{0\mu\nu}^2
+12\,\F D_{\mu}\{D_{\nu}\F,F_{0\mu\nu}\}\right]\,.
\ee
Again, we restrict our attention to static solutions, and as in the $p=1$ case
above we choose
\be
\label{A08}
A_0=\F \implies\ \ D_0\F=0\ \ ,\ \ F_{0\mu}=D_{\mu}\F\ \,,\ \
F_{\mu\nu\rho}=F_{0\mu\nu\rho}\ \ ,\ \ F_{0\mu\nu}=0\,,
\ee
as a result of which the static limit of the Hamiltonian \re{H8} just reduces to the action density of the
$2$-YM system, $i.e.$ only the first term in \re{H8},
whose action is minimised absolutely by the $2$-BPST solution~\cite{Tchrakian:1984gq}.

Using \re{A08}, one can readily verify that the Ampere Law equation \re{ampf8} reduces to
\[
D_{\mu}\{F_{\rho\si},F_{\mu\nu\rho\si}\}=0
\]
which is solved by the selfdual $p=2$ BPST fields, and the Gauss
Law equation \re{gauss8} and Higgs equation \re{higgs8} become identical, reducing to
\be
\label{higgs81}
D_{\mu}\{F_{\rho\si},\{F_{[\rho\si},D_{\mu]}\F\}\}=0\,.
\ee

Up to this point, the $D=4$ and the $D=8$ cases are on the same footing, and hence also is the generic $D=4p$ case.
The remaining task in all these cases is the construction of a regular solution to the Higgs equations \re{higgs41},
\re{higgs81}, $etc$., such that the magnitude of the Higgs field tends to a constant at infinity. This is to guarantee
the convergence of the surface integral
\be
\label{surf}
q\simeq\int_{S^{D-1}}\,\mbox{Tr} \big\{\,\F\,F_{\mu 0} \big\} \,dS^{\mu}\,,
\ee
for the electric flux. The integral \re{surf} will be convergent provided that the Higgs field $\F=A_0$ decays fast enough
at infinity, and, is regular at the origin. Such a solution was found~\cite{Lambert:1999ua} for $D=4$ ($p=1$) but
unfortunately, we have not succeeded to do this for the new, $p\ge 2$ cases. Let us discuss this
question a little further.

In the $D=4$ ($p=1$) case, the Higgs equation \re{higgs41} is solved
most conveniently in the background of the BPST field in the 't~Hooft singular gauge
\be
\label{sing}
A_{\mu}^{(s)}=\frac{1-w(r)}{r}\,\Si_{\mu\nu}\,\hat x_{\nu}\qquad,\qquad w(r)=-\,\frac{\la^2-r^2}{\la^2+r^2}\,,
\ee
by positing the following Ansatz for the Higgs field
\be
\label{ansatzH4}
\F=h(r)\,\Si_{34}\,,
\ee
and noticing~\footnote{Likewise, identifying the prepotential function in the Jackiw-Nohl-Rebbi~\cite{Jackiw:1976fs}
Ansatz with the inverse of the corresponding Higgs field prepotential, yields the multi-dyonic instantons~\cite{Eyras:2000dg}.}
that the inverse of the function multiplying $\Si_{\mu\nu}$ in \re{sing}, namely that
\be
\label{notice1}
h(r)=\frac{r}{1-w(r)}
\ee
solves \re{higgs41}. (In \re{sing}-\re{ansatzH4}, $\Si_{\mu\nu}$ are the chiral representations of the algebra of $SO(4)$.)

The situation with solving \re{higgs81} in the background of \re{sing} (now interpreting $\Si_{\mu\nu}$ there as
the chiral representations of the algebra of $SO(8)$) is a very difficult task, which we have not succeeded in. Even finding an
Ansatz like \re{notice1} for $\F$, which yields a single ordinary differenial equation (ODE) for $h(r)$ is problematic.

Recognising that the Higgs field (or $A_0$) does not necessarily have to take its values inside the algebra of the ``magnetic'' field, we
found an Ansatz for the Higgs field that leads to a single ODE. This is
\be
\label{ansatzH8}
\F=h(r)\,\Si_{5678}\quad,\Si_{5678}=\{\Si_{5[6},\Si_{78]}\}\equiv\{\Si_{56},\Si_{78}\}+{\rm cycl.}\ (678)\,,
\ee
for which the equation \re{higgs81} reduces to one single ordinary differential equation for the function $h(r)$. This is
\be
\label{ode}
-16 \lambda^2(3\lambda^2+2r^2) h
+3r(\lambda^2+r^2) 
\left( (7\lambda^2-r^2)h'
+r(\lambda^2+r^2) h''
\right)=0.
\ee
The asymptotic solutions of \re{ode} are found. As $r\to 0$, $h(r)$ behaves as
\be
\label{g-or}
h(r)= g_2 r^2-\frac{2g_2}{9 \lambda^2}r^4+\frac{4g_2}{27 \lambda^4}r^6+O(r^8)
\ee
(with a free parameter $g_2$), while it decays at infinity as
\be
\label{g-inf}
h(r)=g_0+\frac{4g_0}{ 3r^2}+\frac{34g_0 \lambda^4}{27r^4}+O(1/r^6)
\ee
$g_0$ being another free parameter.

Unfortunately, there exist no solutions $h$ of \re{ode} that satisfy the two asymptotic values in \re{g-or} and \re{g-inf}, simultaneously\footnote{We thank J.~Burzlaff for help on clarifying this point.}.
This can be shown as follows.
Let $ g_2>0$; then according to \re{g-or} $h$ is increasing for small $r$. Choosing $ g_0>0$ in \re{g-inf} for large $r$,
$h'< 0$ in this region, so that $g_2>g_0$. This means that $h$ must have a positive valued maximum, $h_{\rm max}>0$ in between
the two asymptotic regions. At this maximum
$h_{\rm max}'=0$ and $h_{\rm max}''<0$. Substituting $h_{\rm max}>0$, $h_{\rm max}'=0$ and $h''_{\rm max}<0$ in \re{ode},
one sees that the nonvanishing terms there are both
negative, and hence \re{ode} cannot be satified. (This argument can be reversed starting with $ g_2<0$.)

So even with the special Ansatz \re{ansatzH8}, there exist no solutions supporting a nontrivial Higgs field for
the $p\ge 2$ case presented above. Since the Higgs field is necessary for the definition of a (finite) global electric charge,
this means that we cannot find a natural generalization of the LT dyonic instantons
in $d=4p+1>5$ dimensional Minkowski space.

\end{document}